\documentclass[prd, twocolumn, floatfix]{revtex4}
\usepackage{graphicx}
\usepackage{color}
\usepackage{amsmath, amsfonts, amssymb, bm}
\usepackage{slashed}
\begin{document}

\newcommand{\ka}{\slashed{k}}
\newcommand{\A}{\slashed{A}}
\newcommand{\p}{\slashed{p}}
\newcommand{\eps}{\slashed{\varepsilon}}

\title{Dynamically-assisted nonlinear Breit-Wheeler pair production\\
in bichromatic laser fields of circular polarization}
\author{N.~Mahlin, S.~Villalba-Ch\'avez, and C.~M\"uller}
\address{Institut f\"ur Theoretische Physik I, Heinrich-Heine-Universit\"at D\"usseldorf, Universit\"atsstra{\ss}e 1, 40225 D\"usseldorf, Germany}
\date{\today}
\begin{abstract}
Production of electron-positron pairs by a high-energy $\gamma$ photon and a bichromatic
laser wave is considered where the latter is composed of a strong low-frequency 
and a weak high-frequency component, both with circular polarization. An integral expression 
for the production rate is derived that accounts for the strong laser mode to all orders and 
for the weak laser mode to first order. The structure of this formula resembles the well-known 
expression for the nonlinear Breit-Wheeler process in a strong laser field, but includes the 
dynamical assistance from the weak laser mode. We analyze the dependence of the dynamical rate
enhancement on the applied field parameters and show, in particular, that it is substantially 
higher when the two laser modes have opposite helicity.
\end{abstract}

\maketitle

\section{Introduction}
Electron-positron pair production from vacuum by a constant electric field 
is a genuinely nonperturbative process that was first studied by Sauter in 
the early days of relativistic quantum mechanics \cite{Sauter}. Later on, 
Schwinger treated the process within the framework of quantum electrodynamics 
(QED) and established its famous rate $R\sim\exp(-\pi E_c/E_0)$ 
that has been named after him \cite{Schwinger}. It contains the critical field 
of QED, $E_c = m^2c^3/(e\hbar)\approx 1.3\times 10^{16}$\,V/cm, and exhibits a 
non-analytical, manifestly non-perturbative dependence on the applied field 
strength $E_0$. Here, $m$ and $e$ denote the positron mass and charge, respectively. 
Intuitively, the exponential field dependence indicates a
quantum mechanical tunneling from negative-energy to positive-energy states.

Pair production rates in various other strong-field configurations share the
characteristic Schwinger-like form (see \cite{Review1,Review2,Review3,Review4} for reviews). 
For example, pair production in homogeneous electric fields oscillating in 
time \cite{Brezin,Popov} and pair production in combined laser and Coulomb 
fields via the nonlinear Bethe-Heitler effect \cite{Yakovlev, Ritus, Milstein} 
show exponential dependencies on the inverse field strength, as well, provided 
they occur in a quasi-static regime where the pair formation time is much shorter 
than the scale of field variations (and the applied fields are sub-critical).

Because of the huge value of $E_c$, Schwinger pair production and its 
Schwinger-like variants have not been observed experimentally yet \cite{graphene}.
However, motivated by the enormous progress in high-power laser technology,
several high-field laboratories are currently aiming at the detection of 
the fully nonperturbative regime of pair production. They focus, in particular,
on the nonlinear Breit-Wheeler process where pairs are created by a high-energy 
photon colliding with a high-intensity laser wave \cite{Reiss-1962, Nikishov-Ritus, 
Greiner, Heinzl-2010, Krajewska-2012, Meuren-2015, DiPiazza-2016, Blackburn-2018, 
Heinzl-2020}, according to $\omega' + n\omega \to e^+e^-$, with the numbers of 
absorbed laser photons $n\gg 1$. The corresponding rate in the quasi-static regime 
($\xi\gg 1$, $\chi\ll 1$) has the Schwinger-like form $R\sim \exp[-8/(3\chi)]$, where 
$\chi = 2\xi\hbar^2\omega\omega'/(m^2c^4)$ (assuming counterpropagating beams)
denotes the quantum nonlinearity parameter and $\xi = eE_0/(mc\omega)$ is the 
classical laser intensity para\-meter. The experimental realization of this regime---that 
would complement the successful observation of nonlinear Breit-Wheeler pair creation 
in a few-photon regime ($n\sim 5$, $\xi\lesssim 1$) at SLAC in the 1990s \cite{SLAC}---still 
represents a formidable challenge \cite{ELI, E320, CALA, LUXE, RAL}.

To facilitate the observation of Schwinger-like pair production, a mechanism 
termed {\it dynamical assistance} has been proposed theoretically 
\cite{Schutzhold-PRL2008}. It relies on the superposition of a very weak, but 
highly oscillating assisting field onto a strong (quasi)static background. 
Energy absorption from the assisting field can largely enhance the pair 
creation rate, while preserving its nonperturbative character. Dynamically 
assisted pair production has been studied for various field configurations, 
comprising the combination of static and alternating electric fields
\cite{Schutzhold-PRL2008, Orthaber-PLB2011, Grobe-PRA2012, Taya-PRD2019, Selym-PRD2019},
static electric and plane-wave photonic fields \cite{Schutzhold-PRD2009, 
Monin-PRD2010, Torgrimsson-PRD2018} as well as two oscillating electric fields 
with largely different frequencies \cite{Akal-PRD2014, Otto-PLB2015}, 
including spatial field inhomogeneties \cite{Plunien-PRD2018, Schutzhold}. 
Only few studies revealed moreover the impact of dynamical assistance in the 
nonlinear Bethe-Heitler \cite{DiPiazza-PRL2009, Augustin-PLB2014} and 
Breit-Wheeler processes \cite{Jansen-PRA2013}.

In the present paper, we study nonperturbative Breit-Wheeler pair creation 
with dynamical assistance. To this end, the laser field is composed of a 
strong low-frequency and a weak high-frequency component. By 
considering both field components to be circularly polarized and taking
the fermion spins into account, we complement and extend the earlier study 
\cite{Jansen-PRA2013} where dynamically assisted nonlinear Breit-Wheeler 
pair creation of scalar particles has been considered in two mutually 
orthogonal laser field modes of linear polarization. An integral representation 
for the production rate will be derived within the framework of strong-field 
QED, that includes the weak laser mode to leading order and allows to describe 
the absorption of one high-frequency photon from this mode during the pair 
production process. We will show that the latter can lead to a very strong 
dynamical rate enhancement and discuss its dependencies on the applied field 
parameters. In particular, it will be demonstrated that the circular field 
polarization offers an interesting additional setting option because the 
rate enhancement is found to be substantially larger when the two laser modes 
are counter- rather than co-rotating.

It is worth mentioning that, apart from rate enhancements by dynamical assistance,
nonlinear Breit-Wheeler pair production in bichromatic laser fields comprises
further interesting effects. In the case when both field modes have commensurate 
frequencies, characteristic quantum interference effects arise \cite{Yu-PRE1998,
Fofanov-2000, Jansen-2015}, whereas multiphoton threshold effects were presented 
for incommensurate frequencies of similar magnitude \cite{Lyulka}.
A bichromatic field configuration may, moreover, allow for additional pair 
production channels involving photon emission processes \cite{Li-PRD2014}.
And for the so-called laser-assisted Breit-Wheeler process, i.e. pair creation 
in the collision of high-frequency (say, $\gamma$-ray and x-ray) photons 
taking place in the presence of a low-frequency background laser field, 
pronounced redistribution effects in the created particles' phase space have been 
revealed \cite{Nousch-PLB2016}.

Our paper is organized as follows. In Sec.~II we present our analytical approach
to the problem that is based on the $S$ matrix in the Furry picture employing 
Dirac-Volkov states for the fermions. An expression for the rate of Breit-Wheeler 
pair production by absorption of an arbitrary number of photons from the strong 
laser mode and a single photon from the weak laser mode is derived. The physical 
content of this expression is discussed in Sec.~III where we illustrate the rate 
enhancement by dynamical assistance in the nonlinear Breit-Wheeler effect by way
of numerical examples. Our conclusions are given in Sec.~IV. Relativistic units 
with $\hbar=c=4\pi\varepsilon_0=1$ are used throughout, unless explicitly stated 
otherwise. Products of four-vectors are denoted as $(ab) = a_\mu b^\mu = a^0b^0 
- \boldsymbol{a}\cdot\boldsymbol{b}$ and Feynman slash notation is applied. 

\section{Theoretical approach}
In this section we present our analytical treatment of dynamically 
assisted Breit-Wheeler pair production in a bichromatic laser field. 
The latter is described by the four-potential  
\begin{equation}
A_L^\mu(\tau) = A^\mu(\tau) + \tilde{A}^\mu(\tau)
\label{A_L}
\end{equation}
in the radiation gauge and depends on space-time coordinates 
$x^\mu=(t,\boldsymbol{r})$ via the phase variable 
$\tau = (\kappa x) = t - \boldsymbol{\kappa}\cdot\boldsymbol{r}$, 
where $\kappa^\mu = (1,\boldsymbol{\kappa})$ describes the uniform 
wave propagation direction along a unit vector $\boldsymbol{\kappa}$. 
The field is composed of the circularly polarized frequency modes
\begin{eqnarray}
A^\mu(\tau) &=& a\,[ \varepsilon_1^\mu\cos(\eta) + \varepsilon_2^\mu\sin(\eta) ]\ ,\nonumber\\ 
\tilde{A}^\mu(\tau) &=& \tilde{a}\,[ \varepsilon_1^\mu\cos(\tilde{\eta} +
\tilde{\eta}_\alpha) + \sigma\, \varepsilon_2^\mu\sin(\tilde{\eta}+\tilde{\eta}_\alpha) ]
\end{eqnarray}
that will be denoted as main mode and assisting mode, respectively, 
with corresponding frequencies $\omega$ and $\tilde{\omega}$ and wave 
vectors $k^\mu = \omega \kappa^\mu$ and $\tilde{k}^\mu = \tilde{\omega}\kappa^\mu$. 
The phases accordingly read $\eta = (kx) = \omega\tau$, $\tilde{\eta} = 
(\tilde{k}x) = \tilde{\omega}\tau$, whereas $\tilde{\eta}_\alpha$ denotes 
a constant phase shift between the modes. The polarization vectors satisfy 
$(\kappa\varepsilon_i)=0$, $(\varepsilon_i\varepsilon_j)=-\delta_{ij}$ for 
$i,j\in\lbrace1,2\rbrace$. The helicity of the assisting mode is encoded 
in the parameter $\sigma$: the modes are co-rotating for $\sigma=+1$ and 
counter-rotating for $\sigma=-1$. The intensity parameters associated with 
their amplitudes are $\xi=ea/m$ and $\tilde{\xi}=e\tilde{a}/m$.

\subsection{Pair production amplitude}
The $S$ matrix element for nonlinear Breit-Wheeler pair production by a high-energy photon
of wave vector $k'^\mu = (\omega',\boldsymbol{k}')$ and polarization $\varepsilon'^\mu$ in the 
bichromatic laser field \eqref{A_L} reads
\begin{equation} \label{S_fi}
S_\text{fi} = -ie\sqrt{\frac{2\pi}{V\omega^\prime}}\int d^4x\, e^{-i(k^\prime x)}\overline{\Psi}_{p^\prime,s^\prime}^{(-)}\slashed{\varepsilon}^\prime\Psi_{p,s}^{(+)}
\end{equation}
with a normalization volume $V$. Here, $\Psi_{p^\prime,s^\prime}^{(-)}$ and $\Psi_{p,s}^{(+)}$
denote the Volkov states for the created electron, with asymptotic four-momentum $p'^\mu$ 
and spin projection $s'$, and the created positron, with asymptotic four-momentum $p^\mu$ 
and spin projection $s$, respectively. They are given by \cite{Greiner}
\begin{equation} \label{Volkov}
\Psi_{p,s}^{(\pm)}(x) = \sqrt{\frac{m}{V q_L^0}}
\left(1\pm\frac{e\slashed{\kappa}\A_L}{2(\kappa p)}\right)
\begin{Bmatrix} v_{p,s} \\  u_{p,s} \end{Bmatrix}
\exp{\left(iS^{(\pm)}\right)}
\end{equation}
with
$$ S^{(\pm)} = \pm(px)+\frac{e}{(\kappa p)}\int^{\tau}\left[(pA_L(\tau'))
\mp\frac{e}{2}A_L^2(\tau')\right]d\tau'\ .$$
Observe that the normalization constant of the Volkov states is chosen 
with respect to the effective momentum 
\begin{equation} \label{q}
q_L^\mu = p^\mu + \frac{m^2\xi_L^2}{2(\kappa p)}\,\kappa^\mu\ ,
\end{equation}
involving the total intensity parameter $\xi_L=\sqrt{\xi^2+\tilde{\xi}^2}$.

With these details in mind, the $S$ matrix becomes
\begin{equation} \label{S_fi_next}
S_\text{fi} = -ie N_L \int d^4x\, e^{i(q_L^{\prime\mu}+q_L^\mu-k^{\prime\mu})x_\mu}\,\overline{u}_{p^\prime,s^\prime} M_L v_{p,s}\, e^{i\Phi_L}
\end{equation}
with the normalization factor $N_L = \big(\frac{m}{Vq^0_L}\,\frac{m}{Vq_L^{\prime0}}\,\frac{2\pi}{V\omega^\prime}\big)^{1/2}$, the matrix
\begin{equation} \label{M_L}
M_L = \left(1-\frac{e\slashed{A}_L\slashed{\kappa}}{2(\kappa p^\prime)}\right)\slashed{\varepsilon}^\prime\left(1+\frac{e\slashed{\kappa}\slashed{A}_L}{2(\kappa p)}\right) ,
\end{equation}
and the oscillating phase
\begin{eqnarray}\label{Phi_L}
\Phi_L &=& z\,\sin(\eta-\eta_0) + \tilde{z}\,\sin(\tilde{\eta}+\tilde{\eta}_\alpha - \sigma\eta_0) \nonumber\\ & & +\, \tilde{z}_\alpha\,\sin(\tilde{\eta}+\tilde{\eta}_\alpha - \sigma\eta)\, .
\end{eqnarray}
In the latter, we used $z=ea\sqrt{-Q_L^2}$, $\tilde{z}=e\tilde{a}\sqrt{-\tilde{Q}_L^2}$, and
\begin{equation*}
\tilde{z}_\alpha = e^2a\tilde{a}\Bigg[\frac{1}{(\tilde{k}p) -\sigma (kp)}
+ \frac{1}{(\tilde{k}p^\prime) - \sigma(kp^\prime)}\Bigg]
\end{equation*}
with $Q_L^\mu = q_L^\mu/(kq_L) - q_L^{\prime\mu}/(kq_L^\prime)$,
$\tilde{Q}_L^\mu = q_L^\mu/(\tilde{k}q_L) - q_L^{\prime\mu}/(\tilde{k}q_L^\prime)$,
and the angle $\eta_0$ being determined by 
\begin{equation}
\cos(\eta_0) = ea\,\frac{(\varepsilon_1Q_L)}{z}\ ,\ \ \sin(\eta_0) = ea\,\frac{(\varepsilon_2Q_L)}{z}\ .
\end{equation}

By virtue of the Volkov states \eqref{Volkov}, the $S$ matrix \eqref{S_fi} contains both modes of the 
bichromatic laser field to all orders. For our purposes, however, it is possible to simplify
this general expression. Being interested in the Breit-Wheeler process with dynamical assistance,
we shall assume from now on that $\tilde{\xi}\ll 1\lesssim\xi $ and $\omega\ll\tilde{\omega}$.
We may therefore expand the $S$ matrix in powers of the assisting mode amplitude according to \cite{perturbative}
\begin{equation}\label{Taylor}
S_\text{fi} = S_\text{fi}^{(0)} + S_\text{fi}^{(1)} + S_\text{fi}^{(2)} + \ldots
\end{equation}
where $S_\text{fi}^{(j)}\sim\tilde{a}^j$. The zeroth order $S_\text{fi}^{(0)}$ is obtained 
by setting $\tilde{a}=0$ in Eq.~\eqref{S_fi_next}; it coincides with the well-known
expression for the nonlinear Breit-Wheeler process in a monochromatic circularly polarized 
laser wave \cite{Reiss-1962, Nikishov-Ritus, Greiner}
\begin{eqnarray} \label{S_0}
S_\text{fi}^{(0)} &=& -ieN_\text{fi}\!\sum_{n=-\infty}^{\infty}\overline{u}_{p^\prime,s^\prime}\,M_n\,v_{p,s} \nonumber\\
& & \times\,(2\pi)^4\,\delta^4\big(q^{\prime\mu}+q^\mu-k^{\prime\mu}-nk^\mu\big)\,.
\end{eqnarray}
The effective momenta $q^\mu$ and $q^{\prime\mu}$ result from Eq.~\eqref{q} by setting 
$\tilde{a}=0$ therein, i.e., $q^{(\prime)\mu} = p^{(\prime)\mu} + \frac{m^2\xi^2}{2(\kappa p^{(\prime)})}\,\kappa^\mu$; the normalization factor becomes $N_\text{fi} = \big(\frac{m}{Vq^0}\,\frac{m}{Vq^{\prime0}}\,\frac{2\pi}{V\omega^\prime}\big)^{1/2}$, accordingly.
The four-dimensional $\delta$ function displays the energy-momentum conservation in 
the process, and the sum over the number $n$ of photons from the main mode $A^\mu$ originates from a
Fourier series expansion of the periodic parts in the $S$ matrix, according to the formula
$e^{iz\sin(\eta-\eta_0)}=\sum_n J_{-n}(z) e^{-in(\eta-\eta_0)}$ with the ordinary Bessel 
functions $J_n$. The matrix $M_n$ will be given in Eq.~\eqref{M_n} below. Here and in 
the following, the zeroth order contributions are displayed to facilitate a direct comparison 
with terms involving the dynamical assistance by the weak mode $\tilde{A}^\mu$.

\begin{figure}[t]
\begin{center}
\includegraphics[width=0.45\textwidth]{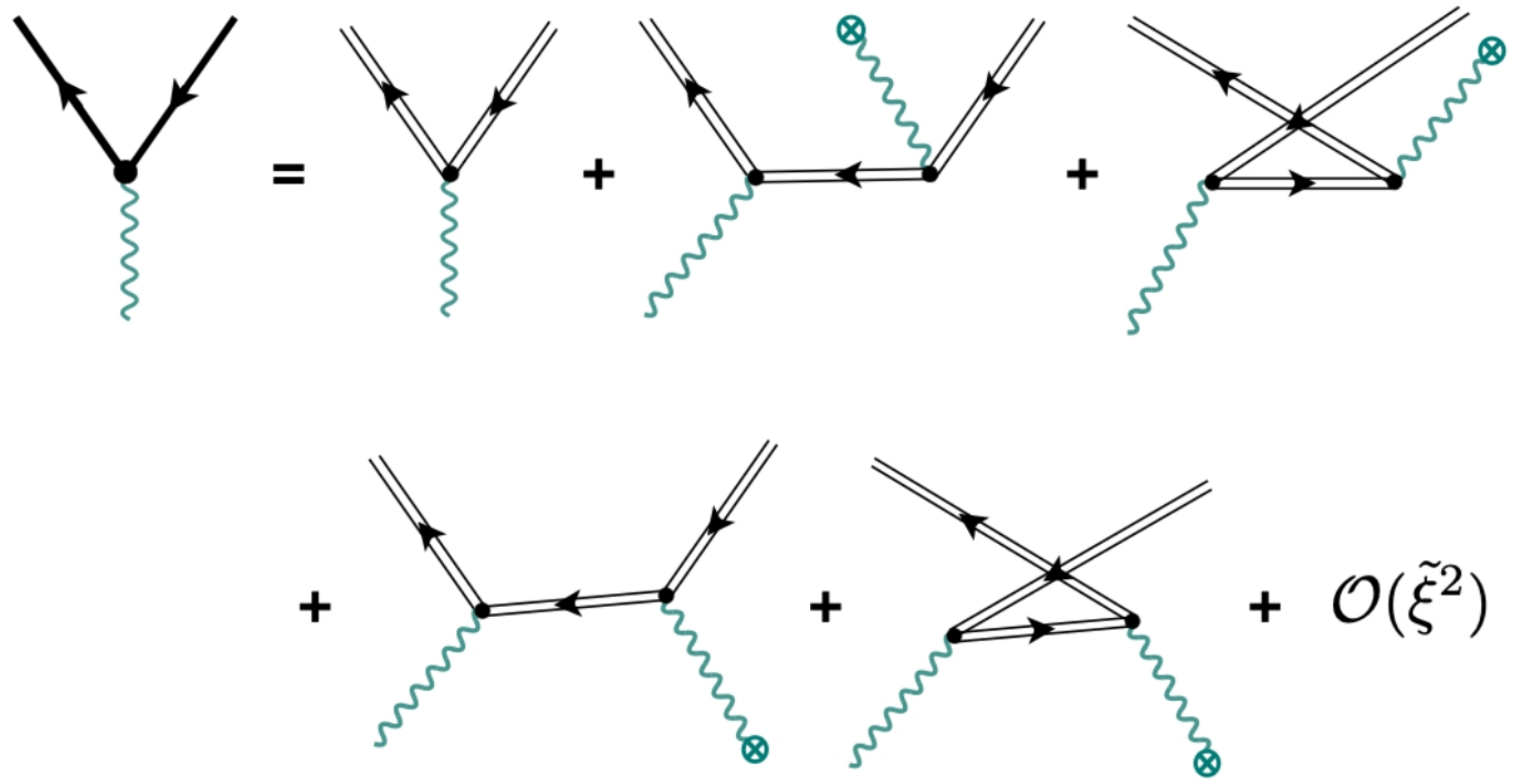}
\caption{Diagrammatic representation of the expansion up to linear order in the assisting mode amplitude of the $S$-matrix element for Breit-Wheeler pair production in the presence of a bichromatic laser field [see Eqs.~\eqref{Taylor}-\eqref{S_1}). The wavy leg, common for all graphs, represents a quantized high-energy photon, whereas those starting (ending) with crossed blobs denote an absorbed (emitted) ``assisting'' photon. The solid  arrowed lines on the left-hand side  stand for the electron and positron wave functions interacting with the bichromatic background wave. Conversely, the external solid double lines on the right-hand side are the corresponding Volkov states which include the interaction with the main mode only. The internal double lines represent the electron-positron propagators in the field of the main mode. Note that, when the amplitude of the main mode vanishes, the leading-order term of the $S$-matrix element reduces to the well-established contributions linked to the linear Breit-Wheeler process.}
\label{fig:diagram}
\end{center}
\end{figure}

The leading order contribution $S_\text{fi}^{(1)}$ is obtained by collecting the 
$\tilde{A}^\mu$ terms from the electronic and positronic Volkov states contained in 
Eq.~\eqref{M_L} as well as the terms linear in $\tilde{z}$ and $\tilde{z}_\alpha$ stemming
from a Taylor expansion of the phase factor $e^{i\Phi_L}$ in Eq.~\eqref{S_fi_next}. The 
resulting expression can be decomposed into two terms,  
$S_\text{fi}^{(1)} = S_\text{fi}^{(1,+)}+S_\text{fi}^{(1,-)}$, with 
\begin{eqnarray} \label{S_1}
S_\text{fi}^{(1,\pm)}\! &=&\! -ieN_\text{fi}\sum_{n=-\infty}^\infty \overline{u}_{p^\prime,s^\prime}\Big(\widetilde{\mathcal{M}}_n^\pm\pm\widetilde{M}_n^\pm\Big)v_{p,s} \nonumber \\
& & \times\,(2\pi)^4\,\delta^4\big(q^{\prime\mu}+q^\mu - k^{\prime\mu} - nk^\mu \pm \tilde{k}^\mu\big)
\end{eqnarray}
corresponding to the emission ($S_\text{fi}^{(1,+)}$) or absorption ($S_\text{fi}^{(1,-)}$) of one photon $\tilde{k}$ from the assisting mode. Our approach is illustrated diagrammatically in Fig.~\ref{fig:diagram}.

The matrices $M_n$, $\widetilde{M}_n^\pm$, $\widetilde{\mathcal{M}}_n^\pm$ in Eqs.~\eqref{S_0} and \eqref{S_1} can be expressed as
\begin{widetext}
\begin{equation}\label{M_n}
\begin{Bmatrix}
M_n\\ \\\widetilde{M}_n^\pm\\ \\\widetilde{\mathcal{M}}_n^\pm
\end{Bmatrix}
=\begin{Bmatrix}
\left(\eps^\prime\!-\!\frac{e^2a^2}{2}\frac{(\varepsilon^\prime \kappa)\slashed{\kappa}}{(\kappa p)(\kappa p^\prime)}\right)B_{-n}\\\vspace{-5pt} \\ \left(\eps^\prime\!-\!\frac{e^2a^2}{2}\frac{(\varepsilon^\prime \kappa)\slashed{\kappa}}{(\kappa p)(\kappa p^\prime)}\right)\widetilde{B}_{-n}^\pm \\ \vspace{-5pt} \\ -\frac{e^2a\tilde{a}}{2}\frac{(\varepsilon^\prime \kappa)\slashed{\kappa}}{(\kappa p)(\kappa p^\prime)}\ \mathcal{B}_{-n}^\pm
\end{Bmatrix}
- \left(\frac{\eps_1\slashed{\kappa}\eps^\prime}{2(\kappa p^\prime)}\!-\!\frac{\eps^\prime\slashed{\kappa}\eps_1}{2(\kappa p)}\right) \begin{Bmatrix}
ea\,C_{-n}\\ \\ ea\,\widetilde{C}_{-n}^\pm\\ \\ e\tilde{a}\,\mathcal{C}_{-n}^\pm
\end{Bmatrix}
- \left(\frac{\eps_2\slashed{\kappa}\eps^\prime}{2(\kappa p^\prime)}\!-\!\frac{\eps^\prime\slashed{\kappa}\eps_2}{2(\kappa p)}\right) \begin{Bmatrix}
ea\,D_{-n}\\ \\ ea\,\widetilde{D}_{-n}^\pm\\ \\ e\tilde{a}\,\mathcal{D}_{-n}^\pm
\end{Bmatrix}
\end{equation}
\end{widetext}
with the known coefficients $B_n=J_n(z)e^{-in\eta_0}$, $C_n = \frac{1}{2}\left(B_{n-1}+B_{n+1}\right)$, and $D_n = \frac{1}{2i}\left(B_{n-1}-B_{n+1}\right)$ in the matrix $M_n$ of the ordinary nonlinear Breit-Wheeler process \cite{Reiss-1962, Nikishov-Ritus, Greiner}. The remaining coefficients 
\begin{eqnarray}\label{B_n-tilde}
\widetilde{B}_n^\pm\! &=&\! \frac{1}{2}\Big[\tilde{z}J_n(z)+\tilde{z}_\alpha 
J_{n\pm\sigma}(z)\Big]\,e^{-i(n\pm\sigma)\eta_0}\, e^{\pm i\tilde{\eta}_\alpha} \nonumber\\
\widetilde{C}_n^\pm\! &=& \!\frac{1}{2}\left(\widetilde{B}_{n-1}^\pm+\widetilde{B}_{n+1}^\pm\right)\,,\
\widetilde{D}_n^\pm =\frac{1}{2i}\left(\widetilde{B}_{n-1}^\pm-\widetilde{B}_{n+1}^\pm\right) \nonumber\\
\mathcal{B}_n^\pm\! &=&\! B_{n\pm\sigma}\,e^{\pm i\tilde{\eta}_\alpha}\,,\
\mathcal{C}_n^\pm = \frac{1}{2}\,\mathcal{B}_{n\mp\sigma}^\pm\,,\
\mathcal{D}_n^\pm = \pm\frac{\sigma}{2i}\,\mathcal{B}_{n\mp\sigma}^\pm \nonumber\\
\end{eqnarray}
in the matrices $\widetilde{M}_n^\pm$ and $\widetilde{\mathcal{M}}_n^\pm$ 
are associated with the first-order contribution $S_\text{fi}^{(1)}$.  
We note that the coefficients $\widetilde{B}^\pm$, $\widetilde{C}^\pm$, and
$\widetilde{D}^\pm$ marked by a tilde vanish in the limit $\tilde{a}\to 0$. 
The coefficients $\mathcal{B}^\pm$, $\mathcal{C}^\pm$, and $\mathcal{D}^\pm$
do not vanish themselves in this limit, but are multiplied by a factor 
$\tilde{a}$ in the matrix $\widetilde{\mathcal{M}}_n^\pm$ [see Eq.~\eqref{M_n}].

All terms in $S_\text{fi}^{(1,\pm)}$ scale linearly with $\tilde{a}$. The 
corresponding contributions to the pair production rate [see Eq.~\eqref{R_1}
below] will therefore contain an additional factor $\tilde{\xi}^2$ as 
compared to the ordinary nonlinear Breit-Wheeler process. In case of 
$S_\text{fi}^{(1,-)}$, the reduction by the factor $\tilde{\xi}^2\ll 1$ is, however,
counteracted by the modified energy-momentum balance: due to the absorption
of one assisting photon $\tilde{k}$, the remaining barrier---that has to be
overcome by additional photon absorption from the main mode---is lowered
which facilitates the pair production. For suitably chosen field parameters,
the latter effect can overcompensate the reduction from the $\tilde{\xi}^2$
scaling, this way leading to enhanced pair production. This is the physical
origin of the rate enhancement by dynamical assistance. Conversely, the other 
first-order term $S_\text{fi}^{(1,+)}$, involving the emission of a photon
$\tilde{k}$ into the assisting mode, corresponds to an even higher barrier
that has to be overcome by photon absorption from the main laser mode. It
will give a smaller contribution to the rate than the assistance-free term 
$S_\text{fi}^{(0)}$ and does not play a role for the enhancement effect that we aim for.

\subsection{Pair production rate}
From the $S$ matrix we obtain the production rate per incident $\gamma$ photon
by taking the absolute square, summing over the produced particle spins, 
intergrating over their momenta, averaging over the $\gamma$-photon polarizations, 
and dividing out the interaction time: 
\begin{equation}\label{R}
R = \int \frac{1}{2}\sum_{\lambda'}\sum_{s,s'}\frac{|S_\text{fi}|^2}{T}\,\frac{V d^3q}{(2\pi)^3}\,\frac{V d^3q^\prime}{(2\pi)^3}\,.
\end{equation}
The production rate $R$ accordingly refers to a scenario where the incident beam 
of $\gamma$ photons is unpolarized.

In general, the absolute square of $S_\text{fi}$ from Eq.~\eqref{Taylor} 
contains---apart from diagonal terms---also cross terms that describe
interferences between different contributions. In particular, the 
cross term of $S_\text{fi}^{(0)}$ with $S_\text{fi}^{(1)}$ leads to
rate contributions linear in $\tilde{a}$. These terms would exist and
could cause interesting effects if the ratio of $\tilde{\omega}$
and $\omega$ was an integer (e.g. $\tilde{\omega}=2\omega$). Such two-color 
quantum interference effects have already been studied elsewhere
\cite{Yu-PRE1998, Fofanov-2000, Jansen-2015}; they are not of interest
in the current consideration. 
To be specific, we shall assume in the following that the frequency
ratio $\tilde{\omega}/\omega$ is not an integer. Then the cross terms
between $S_\text{fi}^{(0)}$ and $S_\text{fi}^{(1)}$ vanish identically
because the associated $\delta$ functions in Eqs.~\eqref{S_0} and
\eqref{S_1} cannot be satisfied simultaneously. By requiring more 
strictly that $2\tilde{\omega}/\omega$ is not an integer either, we
can moreover exclude interferences between $S_\text{fi}^{(1,+)}$ and
$S_\text{fi}^{(1,-)}$ as well as between $S_\text{fi}^{(0)}$ and 
those terms in $S_\text{fi}^{(2)}$ which describe the absorption 
(or emission) of two photons $\tilde{k}$ \cite{Jansen-PRA2013}.
And by finally imposing the stricter condition
\begin{equation}\label{F2}
6\tilde{\omega}/\omega\notin\mathbb{N}
\end{equation}
also interference terms of order $\tilde{a}^3$ between
$S_\text{fi}^{(0)}$ and $S_\text{fi}^{(3)}$ as well as between 
$S_\text{fi}^{(1)}$ and $S_\text{fi}^{(2)}$ drop out.

Under this assumption, the squared $S$ matrix becomes
\begin{equation}\label{S_square}
|S_\text{fi}|^2 = \big\lvert S_\text{fi}^{(0)}\big\rvert^2_\sim+\big\lvert S_\text{fi}^{(1,+)}\big\rvert^2+\big\lvert S_\text{fi}^{(1,-)}\big\rvert^2+\mathcal{O}(\tilde{a}^4)
\end{equation}
with the terms $\big\lvert S_\text{fi}^{(1,\pm)}\big\rvert^2$ being of order $\tilde{a}^2$. We note
that the first term $\big\lvert S_\text{fi}^{(0)}\big\rvert^2_\sim$ on the right-hand side of Eq.~\eqref{S_square} comprises the $\tilde{a}$-independent contribution $\lvert S_\text{fi}^{(0)}\rvert^2$ along with $\mathcal{O}(\tilde{a}^2)$ corrections to it. They stem from interferences between $S_\text{fi}^{(0)}$ and second-order terms in $S_\text{fi}^{(2)}$ associated with the simultaneous absorption and emission of an assisting photon $\tilde{k}$ \cite{q_L}. Leading to the same energy-momentum balance as in Eq.~\eqref{S_0}, these emission-absorption processes are strongly suppressed as compared with $S_\text{fi}^{(0)}$, since they scale with $\tilde{\xi}^2\ll 1$ but do not lower the pair production barrier to be overcome by photon absorption from the main mode. They can therefore be safely neglected. In contrast, the dynamical assistance described by $\big\lvert S_\text{fi}^{(1,-)}\big\rvert^2$ can largely dominate over the $\big\lvert S_\text{fi}^{(0)}\big\rvert^2$ term, as will be demonstrated by numerical examples in Sec.~III.

The spin summation and polarization average in Eq.~\eqref{R} can be carried out
in the usual way by taking traces over the involved Dirac $\gamma$ matrices.
The result for the ordinary nonlinear Breit-Wheeler process is
\begin{eqnarray}\label{M_fi_0}
& &\frac{1}{2}\sum_{\lambda^\prime}\sum_{s,s^\prime}\,|\overline{u}_{p^\prime,s^\prime}M_n v_{p,s}|^2 \nonumber\\
& & = |B_{-n}|^2 + \xi^2 \bigg[\left(|B_{-n}|^2 - |C_{-n}|^2 - |D_{-n}|^2\right) \nonumber\\
& & \times \left(1-\frac{(\kappa k^\prime)^2}{2(\kappa p)(\kappa p^\prime)} \right)\bigg],
\end{eqnarray}
whereas for the leading-order in $\tilde{a}$ terms we obtain
\begin{eqnarray}\label{M_fi_pm}
& & \frac{1}{2}\sum_{\lambda^\prime}\sum_{s,s^\prime}\big\lvert\overline{u}_{p^\prime,s^\prime}\Big(\widetilde{\mathcal{M}}_n^\pm\pm\widetilde{M}_n^\pm\Big)v_{p,s}\big\rvert^2\nonumber\\
& & = \big\lvert\widetilde{B}_{-n}^\pm\big\rvert^2+\xi^2\bigg[\bigg(\big\lvert\widetilde{B}_{-n}^\pm\big\rvert^2\pm\mathfrak{Re}\!\left[\mathcal{B}_{-n}^\pm\!\left(\widetilde{B}_{-n}^\pm\right)^{\!\ast}\right]\frac{\tilde{a}}{a}\nonumber\\
& &\ \ -\Big\lvert\mathcal{C}_{-n}^\pm\,\frac{\tilde{a}}{a}\pm\widetilde{C}_{-n}^\pm\Big\rvert^2-\Big\lvert\mathcal{D}_{-n}^\pm\,\frac{\tilde{a}}{a}\pm\widetilde{D}_{-n}^\pm\Big\rvert^2\bigg)\nonumber\\
& &\ \ \times \bigg(1-\frac{(\kappa k^\prime)^2}{2(\kappa p)(\kappa p^\prime)}\bigg)\bigg].
\end{eqnarray}
By performing afterwards the integrations over the particle momenta 
$\boldsymbol{q}$ and $\boldsymbol{q}'$, we obtain the corresponding 
contributions to the pair production rate, 
\begin{equation}\label{R_sum}
R=R^{(0)}+R^{(1,+)}+R^{(1,-)}+\mathcal{O}(\tilde{\xi}^{\,4})\,.
\end{equation}
The well-established zeroth order contribution reads \cite{Reiss-1962, Nikishov-Ritus, Greiner}
\begin{eqnarray}\label{R_0}
R^{(0)}\! &=&\! \frac{\alpha\,m^2}{4\omega^\prime}\sum_{n\geq n_0}^\infty\int_1^{u_n}\!\!\!\frac{d u}{u\sqrt{u(u-1)}}\big[2J_n^2+\xi^2 \nonumber\\
&\times&\! \big(J_{n+1}^2+J_{n-1}^2-2J_n^2\big)(2u-1)+\mathcal{O}(\tilde{\xi}^{\,2})\big],
\end{eqnarray}
where correction terms of order $\tilde{a}^2$ from combined emission-absorption 
photon exchange processes with the assisting mode that do not change the four-momentum 
balance have been neglected. Here, $\alpha = e^2$ is the fine-structure constant and 
$n_0=4m_\ast^2/s$ the photon number threshold, with the effective fermion mass 
$m_\ast=m\sqrt{1+\xi^2}$ dressed by the main laser mode and the Mandelstam variable 
$s=2(kk^\prime)$. Besides, the upper integration limit is $u_n=n/n_0$, and the Bessel 
functions $J_\nu=J_\nu(z)$ depend on the argument $z=(8m^2/s)\,\xi\sqrt{1+\xi^2}\sqrt{u(u_n-u)}$.

For the leading-order contribution with respect to $\tilde{a}$, which involves
the absorption or emission of one photon $\tilde{k}$ from the assisting mode,
we find
\begin{widetext}
\begin{align}\label{R_1}
R^{(1,\pm)}=&\,\frac{\alpha\,m^2}{4\omega^\prime}\sum_{n\geq n_0^\pm}^\infty\int_1^{u_{\tilde{n}^\pm}}\!\!\!\frac{d u}{u\sqrt{u(u-1)}}\Bigg[\frac{1}{2}\bigg(\tilde{z}^{\pm\,2}J_n^2+\tilde{z}_\alpha^2J_{n\mp\sigma}^2-2\tilde{z}^\pm\tilde{z}_\alpha J_nJ_{n\mp\sigma}\bigg)\nonumber\\
&+\frac{1}{4}\bigg(\xi^2\Big[\tilde{z}^{\pm\,2}\big(J_{n+1}^2+J_{n-1}^2-2J_n^2\big)+\tilde{z}_\alpha^2\big(J_n^2+J_{n\mp 2\sigma}^2-2J_{n\mp\sigma}^2\big) - 2\tilde{z}^\pm\tilde{z}_\alpha\big(J_{n\pm\sigma}J_n+J_{n\mp\sigma}J_{n\mp 2\sigma}-2J_nJ_{n\mp\sigma}\big)\Big] \nonumber \\
& - 4\xi\tilde{\xi}\Big[\tilde{z}^\pm J_n \big(J_{n+\sigma} - J_{n-\sigma}\big)\mp \tilde{z}_\alpha\big(J_n^2-J_{n\mp\sigma}^2\big)\Big] + 4\tilde{\xi}^{\,2}J_n^2\bigg)(2u-1)\Bigg]
\end{align}
\end{widetext}
with 
\begin{eqnarray}
n_0^\pm\! &=&\! n_0\pm \frac{\tilde{s}}{s}\,,\ u_{\tilde{n}^\pm}=\frac{ns\mp\tilde{s}}{n_0^\pm s\mp\tilde{s}}\,,\nonumber\\
\tilde{s}\! &=&\! 2(\tilde{k}k^\prime)\,,\ 
\tilde{z}_\alpha = \,\frac{8m^2}{\tilde{s} - \sigma s}\,\xi\tilde{\xi}\,u\,, \nonumber\\
\tilde{z}^\pm\! &=&\! \frac{8m^2}{\tilde{s}}\,\tilde{\xi}\sqrt{1+\xi^2}\sqrt{u(u_{\tilde{n}^\pm}-u)}
\end{eqnarray}
and the Bessel functions $J_\nu=J_\nu(z^\pm)$ depending on the argument
$z^\pm=(8m^2/s)\,\xi\sqrt{1+\xi^2}\sqrt{u(u_{\tilde{n}^\pm}-u)}$.

Equation~\eqref{R_1} constitutes the main result of our paper.
While being somewhat more involved, its general structure---containing an 
integral over the variable $u$, that is related to the polar emission angle 
of the created particles, and a sum over the number of photons absorbed 
from the strong main laser mode---closely resembles the rate expression 
\eqref{R_0} for the ordinary nonlinear Breit-Wheeler process. The important 
difference is that our formula for $R^{(1,-)}$ (or $R^{(1,+)}$) accounts 
for the absorption (or emission \cite{Li-PRD2014}) of an additional 
photon from the assisting weak laser mode. Accordingly, $R^{(1,-)}$
describes nonlinear Breit-Wheeler pair production in circularly polarized
laser fields proceeding via the absorption of many low-frequency photons 
from the main field mode and a single high-frequency photon from the
assisting mode, which may have either equal or opposite helicity as the
main mode. The dynamical assistance provided by the weak high-frequency mode can largely
enhance the pair production rate $R^{(1,-)}$ as compared with $R^{(0)}$.

Before moving on to the next section we note that, in the limit when 
the main laser mode vanishes ($\xi\to 0$) while the assisting laser mode
has low amplitude ($\tilde{\xi}\ll 1$) and sufficiently high 
frequency (such that $\tilde{s} > 4m^2$), the expression for $R^{(1,-)}$
reproduces the rate for the original Breit-Wheeler process \cite{Breit-Wheeler} 
of pair production by two photons (see also Fig.~\ref{fig:diagram}).

\section{Numerical results and Discussion}
In this section we illustrate our findings on the dynamically assisted nonlinear 
Breit-Wheeler process in a bichromatic laser field by numerical examples. 
For reasons of computational feasibility the 
main mode intensity parameter is chosen to have a rather moderate value of $\xi\sim 1$,
while the associated frequency is $\omega\approx 0.05m$. The frequency of the 
$\gamma$-beam is taken throughout as $\omega' = 0.706m$. For comparison we note that 
the typical values in experiment are $\omega\sim 1$\,eV and $\omega'\sim 10^{10}$\,eV 
\cite{SLAC, ELI, E320, CALA, LUXE, RAL}, yielding a product of $\omega\omega'\sim 0.04m^2$.
The latter value is closely met by our set of parameters, meaning that we perform our
calculations in a frame of reference that is boosted with respect to the laboratory frame.
The assisting mode parameters are taken as $\tilde{\xi}\sim 10^{-3}$ and 
$\tilde{\omega}\approx \omega'$, describing accordingly a weak mode of high frequency.
The $\gamma$-beam and bichromatic laser wave are assumed to be counterpropagating.

In the following figures we present the rate $R_{\rm DA} := R^{(0)} + R^{(1,+)} + R^{(1,-)}$ 
from Eqs.~\eqref{R_sum}--\eqref{R_1} for nonlinear Breit-Wheeler pair creation in a 
bichromatic laser field, including the effect of dynamical assistance.
Two variants of this rate exist, depending on whether the two laser modes have
equal or opposite helicities. In the figures, the corresponding rates are denoted 
suggestively as $R_{\rm DA}(\sigma)$, with $\sigma=+1$ and $\sigma=-1$ respectively.
Comparing them allows us to reveal the influence of the wave helicities on the 
exerted dynamical assistance. 

The rates $R_{\rm DA}(\sigma)$ in a bichromatic 
laser field will moreover be compared with the corresponding 'monochromatic rates' 
when only one of the two laser modes is present, in order to quantify the enhancement 
effect. This is, first of all, the rate $R^{(0)}$ for the ordinary
nonlinear Breit-Wheeler process where the assisting mode is absent ($\tilde{\xi}=0$);
it is denoted as $R_{\rm BW}$ in the figures. Besides, the rate for nonlinear Breit-Wheeler
pair creation by the $\gamma$-beam and the assisting mode alone, when the strong field
is switched off ($\xi=0$) forms a second reference, denoted by $\tilde{R}_{\rm BW}$.
We note that for the chosen parameters, at least three $\tilde{\omega}$-photons need 
to be absorbed in the latter scenario to overcome the pair creation threshold.

\subsection{Enhanced pair creation by dynamical assistance}

\begin{figure}[t]
\begin{center}
\includegraphics[width=0.52\textwidth]{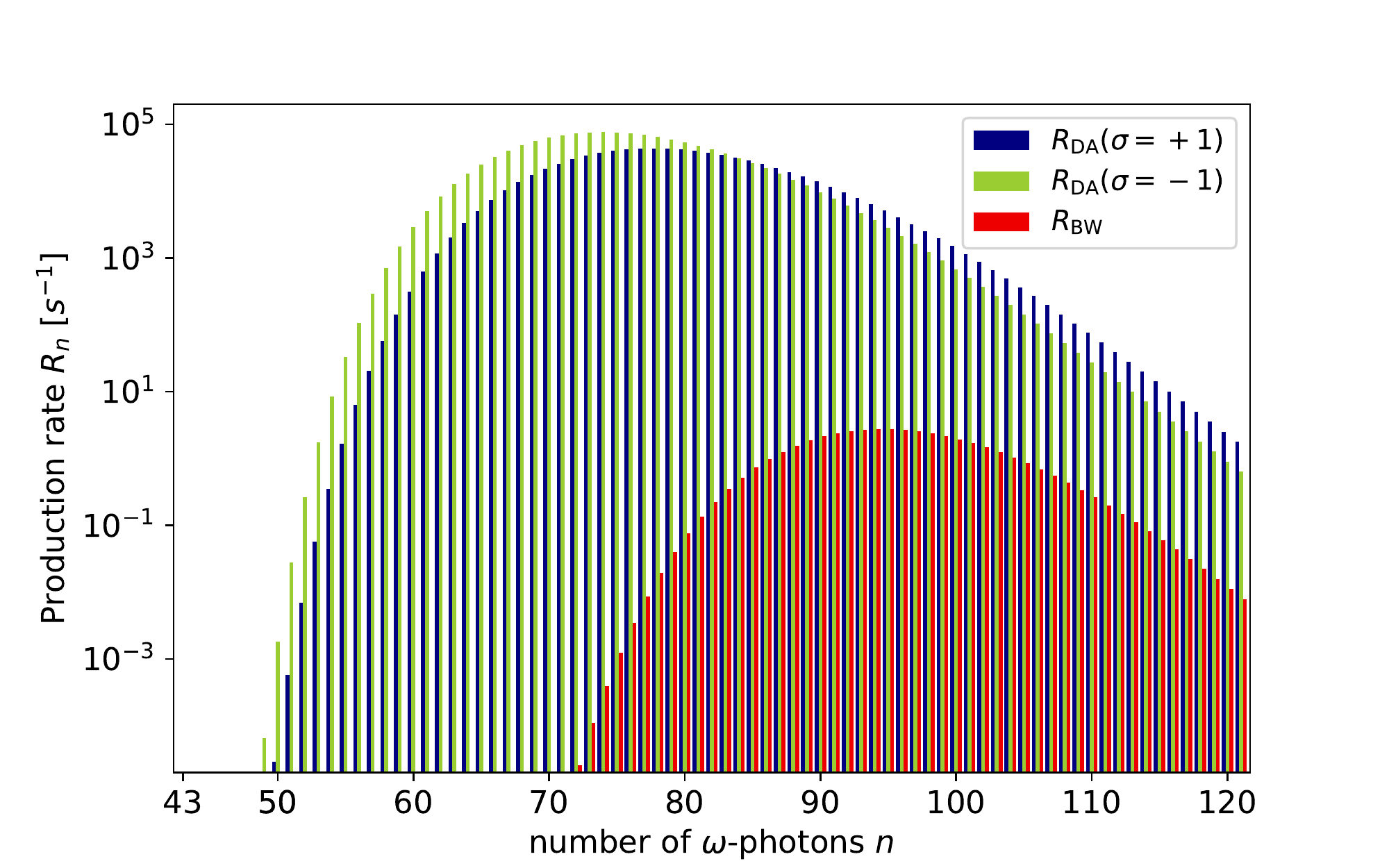}
\caption{Contributions to the pair creation rate in dependence on the number of photons 
absorbed from the main laser mode. The blue (dark gray) and green (light gray) bars 
refer to a bichromatic laser wave with co-rotating and counter-rotating modes, respectively, 
whereas the red (gray) bars show the unassisted case, as indicated in the legend.
The parameters are $\xi = 1$, $\tilde{\xi} = 10^{-3}$, $\omega = 0.05 m$, and
$\tilde{\omega} = \omega' = 0.706 m$.}
\label{fig:number}
\end{center}
\end{figure}

Figure~\ref{fig:number} shows the contributions to the pair creation rate stemming from 
the absorption of $n$ laser photons from the main mode. Already
here a pronounced rate enhancement through the dynamical assistance by the weak laser mode
becomes apparent, as the contributions to $R_{\rm DA}(\sigma)$ are much larger than those to 
$R_{\rm BW}$. They are shifted besides to smaller photon numbers because a part of the
four-momentum required for pair creation already comes from the absorption of the high-frequency 
photon from the weak mode. Note that the photon number distributions in Fig.~\ref{fig:number} 
are shifted to values substantially higher than the photon number thresholds of 
$n_0\approx 56.7$ and $n_0^-\approx 42.5$ [see below Eqs.~\eqref{R_0} and \eqref{R_1}], 
respectively, which is a characteristic above-threshold feature of pair creation at 
$\xi\gtrsim 1$. Furthermore, a comparison of $R_{\rm DA}(\sigma=+1)$ and $R_{\rm DA}(\sigma=-1)$ 
shows an impact of the mode helicities: the number distribution for counter-rotating 
laser modes reaches larger maximum values and is slightly shifted to the left.  

\begin{figure}[t]
\begin{center}
\includegraphics[width=0.5\textwidth]{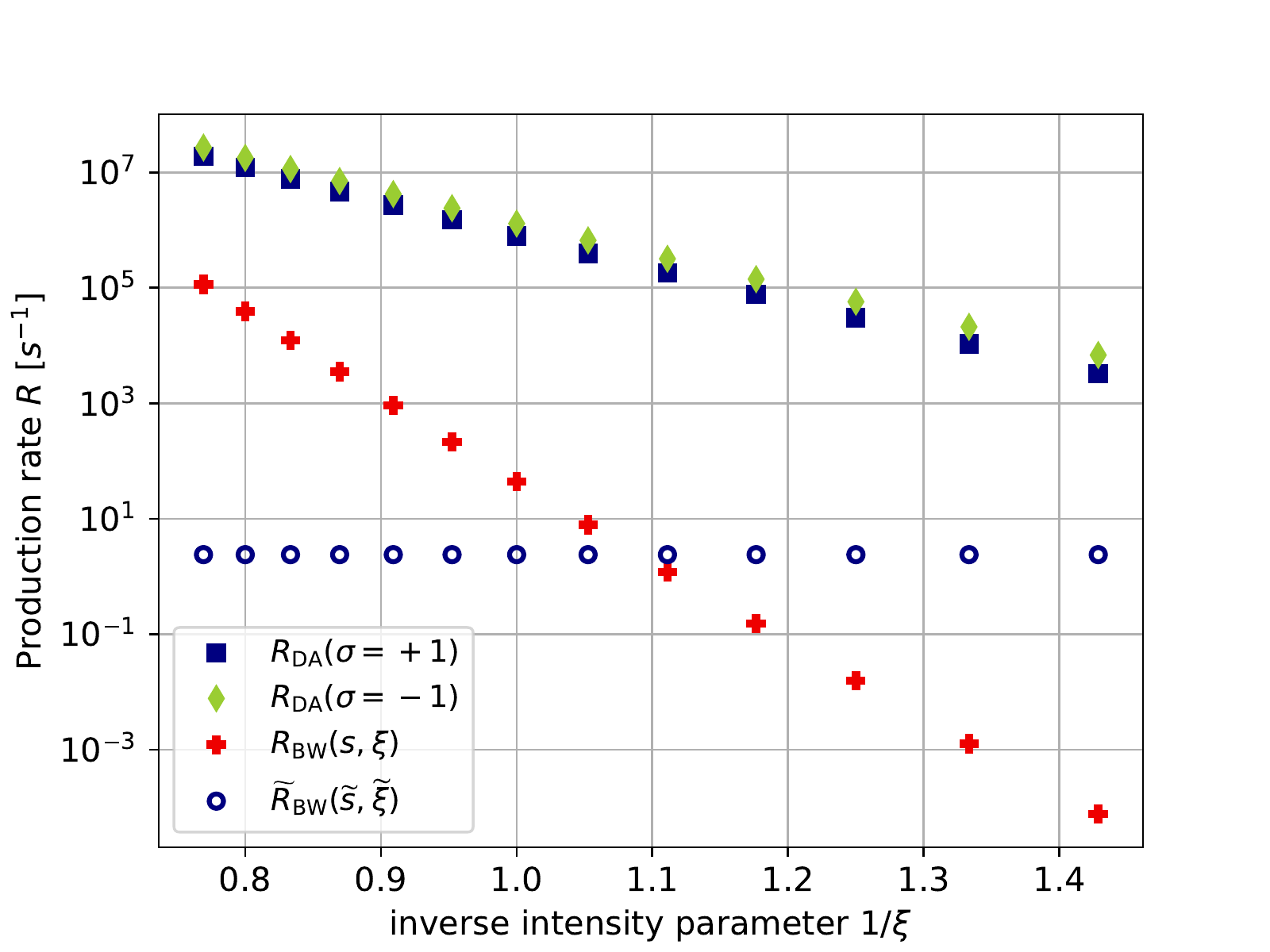}
\caption{Total rates for dynamically assisted Breit-Wheeler pair creation in a bichromatic 
laser wave, as function of the inverse intensity parameter of the main mode; blue squares (green diamonds) refer to co-rotating (counter-rotating) laser modes.
The parameters are the same as in Fig.~\ref{fig:number}. The unassisted case ($\tilde{\xi}=0$) 
is displayed by red crosses. Blue circles show the rate when instead the main mode is absent ($\xi=0$).}
\label{fig:inverse-xi}
\end{center}
\end{figure}

By summing the rate contributions over the number of absorbed strong-field photons,
the corresponding total pair creation rates are obtained. They are shown in Fig.~\ref{fig:inverse-xi},
as function of the inverse value of the main mode intensity parameter. In the chosen 
logarithmic representation, the dynamically assisted rates $R_{\rm DA}(\sigma=\pm 1)$ and
the unassisted rate $R_{\rm BW}$ follow to a very good approximation declining straight
lines, illustrating their Schwinger-like exponential dependence on $-b/\xi$, with some
process-specific parameter $b$. In addition, the rate $\tilde{R}_{\rm BW}$ for pair 
creation by $\gamma$-beam and assisting mode is included for reference along a horizontal line.

In the chosen range of parameters, the rates for dynamically assisted nonlinear Breit-Wheeler 
pair creation lie far above the monochromatic rates. The relative enhancement is largest 
close to $1/\xi\approx 1.1$ (i.e.~$\xi\approx 0.9$) and amounts to almost five orders of magnitude. 
The decline of the curves for assisted pair creation is much slower than for the unassisted
process: by fitting our data to a Schwinger-like exponential, we obtain $b_0\approx 35.0$
for the unassisted process \cite{asymptotic}, while $b_1\approx 14.0$ (13.3) for the assisted 
process with $\sigma=+1$ ($\sigma=-1$). The reduced slope 
arises because the absorption of the high-frequency photon from the weak mode
reduces the tunneling barrier that remains to be overcome \cite{Schutzhold-PRL2008, 
DiPiazza-PRL2009, Augustin-PLB2014, Jansen-PRA2013}.

Figure~\ref{fig:inverse-xi} shows moreover that the effect of dynamical assistance 
is stronger when the two laser modes have opposite helicity. This is remarkable
because (i) the intuitive picture of dynamical assistance relies on the fact that 
the additional energy absorbed from the assisting field facilitates to overcome the 
pair creation threshold and (ii) the four-momentum of the assisting laser photon is 
the same for $\sigma=\pm 1$. Hence, the reduction of the tunneling energy-barrier by 
absorption of a high-frequency $\tilde{\omega}$-photon occurs independently of its helicity.

The more pronounced enhancement effect of an assisting mode with opposite helicity
can be explained by angular momentum conservation. The circularly polarized laser 
photons in our scenario carry definite angular momentum along the propagation axis
of $+\hbar$ or $-\hbar$. When the modes co-rotate, the angular momentum of each 
absorbed photon points in the same direction. In contrast, when the modes counter-rotate, 
the absorption of the assisting photon reduces the total angular momentum of all
absorbed photons. Accordingly, for fixed strong-field photon number $n$, the angular 
momentum that is transferred to the created pair amounts to $(n+\sigma)\hbar$.

In a semiclassical picture, the orbital angular momentum of the electron 
and positron is determined by their relative momentum. In case of the unassisted 
Breit-Wheeler process, it amounts to 
$\ell \approx 2[n_0(n-n_0)]^{1/2}\hbar$ \cite{Ritus-Review}. This quantitiy 
is bounded from above according to $\ell\le n\hbar$. Large contributions 
to the production rate can be expected when $\ell$ approximately balances the 
angular momentum of the absorbed photons \cite{Ritus-Review}. 

In case of the assisted Breit-Wheeler process, we have to demand 
$\ell \approx (n+\sigma)\hbar$ with $\ell \approx 2[n_0^-(n-n_0^-)]^{1/2}\hbar$, 
accordingly. For counter-rotating waves, the condition $\ell = (n-1)\hbar$ is 
very well met for $n\approx 73$ in Fig.~\ref{fig:number} and, indeed, exactly in 
this region of photon numbers we find the highest rate contributions. For co-rotating 
waves, however, the condition $\ell = (n+1)\hbar$ cannot be satisfied because 
$\ell$ is at most $n\hbar$. In this case, the angular momentum balance can only 
be fullfilled when both particle spins are oriented along the laser propagation
direction, this way providing an extra contribution of one unit of $\hbar$ to 
the total angular momentum of the pair. Such an additional constraint is 
absent for counter-rotating waves, so that the accessible spin space is larger 
in this case. As a result, the total pair production rate is higher when the 
modes counter-rotate \cite{semiclassical}. 

Our semiclassical consideration also explains the horizontal shift between the 
number distributions for $\sigma=-1$ and $\sigma=+1$ in Fig.~\ref{fig:number}. 
Since in the latter case, $\ell\approx n\hbar$ is favorable, the region of highest 
rate contributions moves to larger $n$ values than in the former case.
We note that, asymptotically for $\xi\gg 1$, one would expect the highest rates 
at $n\approx 2n_0^-\approx 85$ for co-rotating waves \cite{Ritus-Review}.

\begin{figure}[t]
\begin{center}
\includegraphics[width=0.5\textwidth]{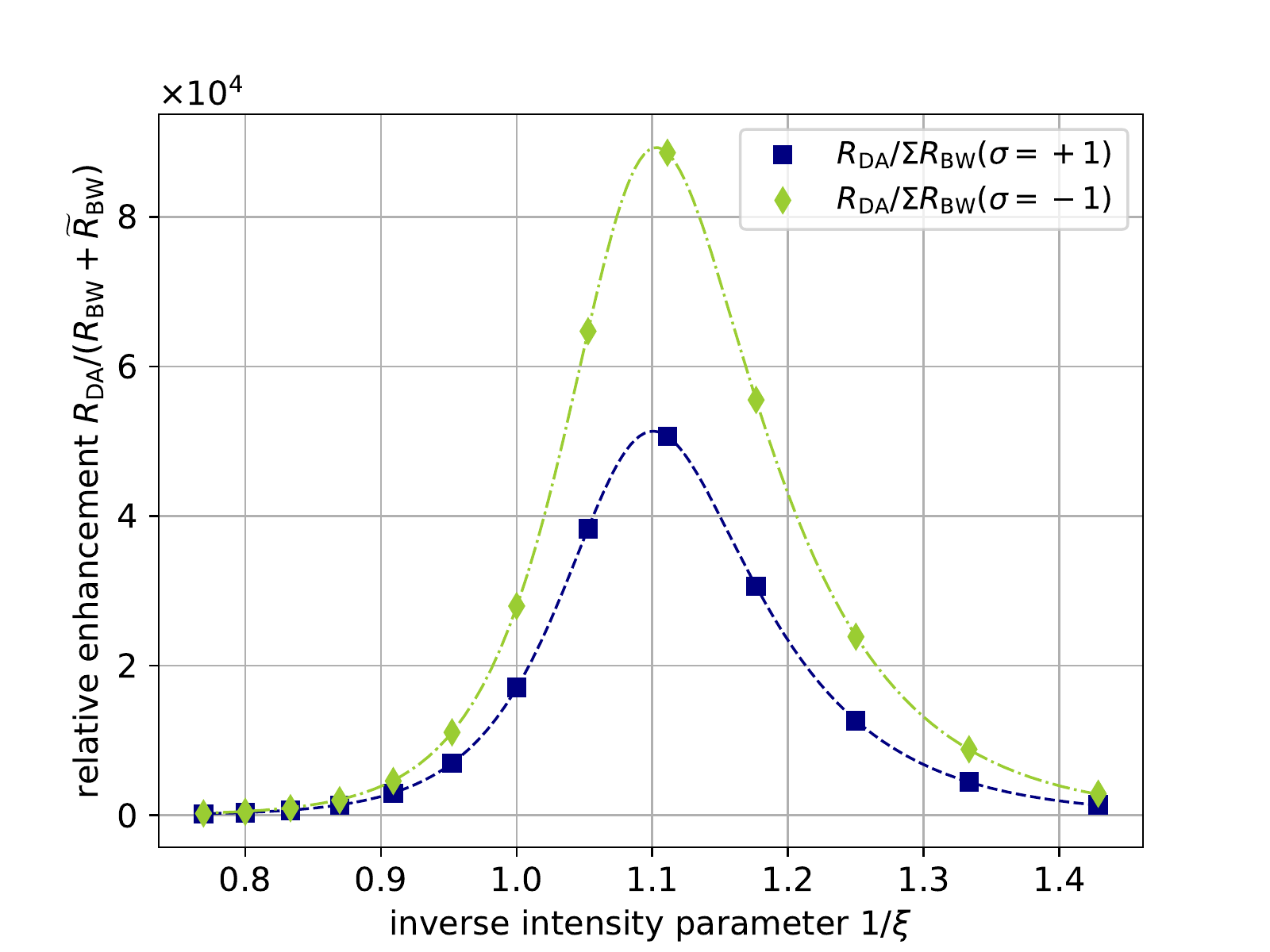}
\caption{Relative enhancement of the dynamically assisted Breit-Wheeler process in a bichromatic
laser field over the sum of the respective monochromatic rates. Blue squares (green diamonds) refer to co-rotating (counter-rotating) laser modes.
The parameters are the same as in Fig.~\ref{fig:number}.}
\label{fig:enhancement}
\end{center}
\end{figure}

The larger effectiveness of an assisting photon of opposite helicity is also seen in 
Fig.~\ref{fig:enhancement}, displaying the relative enhancement due to dynamical
assistance as compared with the sum of the monochromatic rates. At $\xi\approx 0.9$,
the pair creation rate for counter-rotating modes is almost twice as large as for 
co-rotating modes. The bell-shaped form of the relative enhancement curves is a 
consequence of the rate dependencies shown in Fig.~\ref{fig:inverse-xi}. The curves 
reach their maximum close to the point where the monochromatic rates $R_{\rm BW}$ and 
$\tilde{R}_{BW}$ cross. For smaller values of $1/\xi$, the relative enhancement is
reduced due to the different slopes of the dynamically assisted rates $R_{\rm DA}(\sigma)$ 
as compared with the unassisted rate $R_{\rm BW}$. For larger values of $1/\xi$, it 
is reduced as well, since the rates $R_{\rm DA}(\sigma)$ are falling while $\tilde{R}_{\rm BW}$ is 
constant.

The dashed lines in Fig.~\ref{fig:enhancement} show fit functions of the form 
$[c_1\exp(-b_1/\xi) + \exp(-b_0/\xi)] / [c_0 + \exp(-b_0/\xi)]$, with four 
free parameters $b_0$, $b_1$, $c_0$ and $c_1$ for $\sigma=\pm 1$ each.
While in the first place serving to guide the eye, these fit functions have a
physically motivated form, relying on the Schwinger-like exponential $\xi$-dependencies
of the assisted and unassisted nonlinear Breit-Wheeler rates and the 
$\xi$-independency of $\tilde{R}_{\rm BW}$. One obtains here $b_0\approx 32.4$ 
and $b_1\approx 12.7$ for $\sigma=+1$, whereas $b_1\approx 12.1$ for $\sigma=-1$.

\subsection{Parameter study of relative enhancement}
For the field parameters considered in the previous section, we found a relative 
enhancement of the dynamically assisted rates $R_{\rm DA}(\sigma)$ over the sum
of the monochromatic rates $R_{\rm BW}$ and $\tilde{R}_{\rm BW}$ up to about 
$5\times 10^4$ for $\sigma=+1$ and $9\times 10^4$ for $\sigma=-1$ (see 
Fig.~\ref{fig:enhancement}). In the following we will study the dependencies
of the relative enhancement due to dynamical assistance on the applied parameters.

\begin{figure}[t]
\begin{center}
\includegraphics[width=0.52\textwidth]{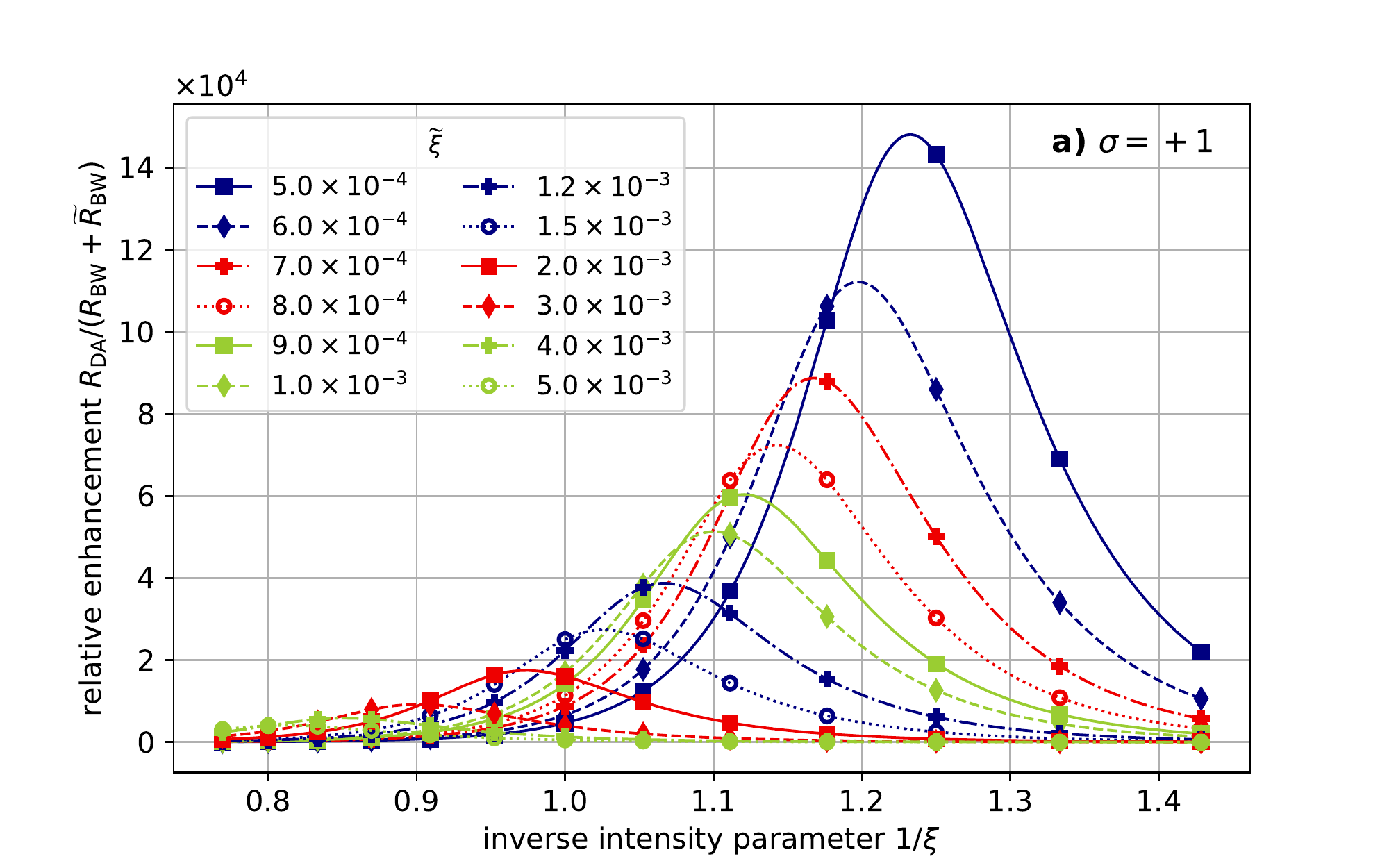}
\includegraphics[width=0.52\textwidth]{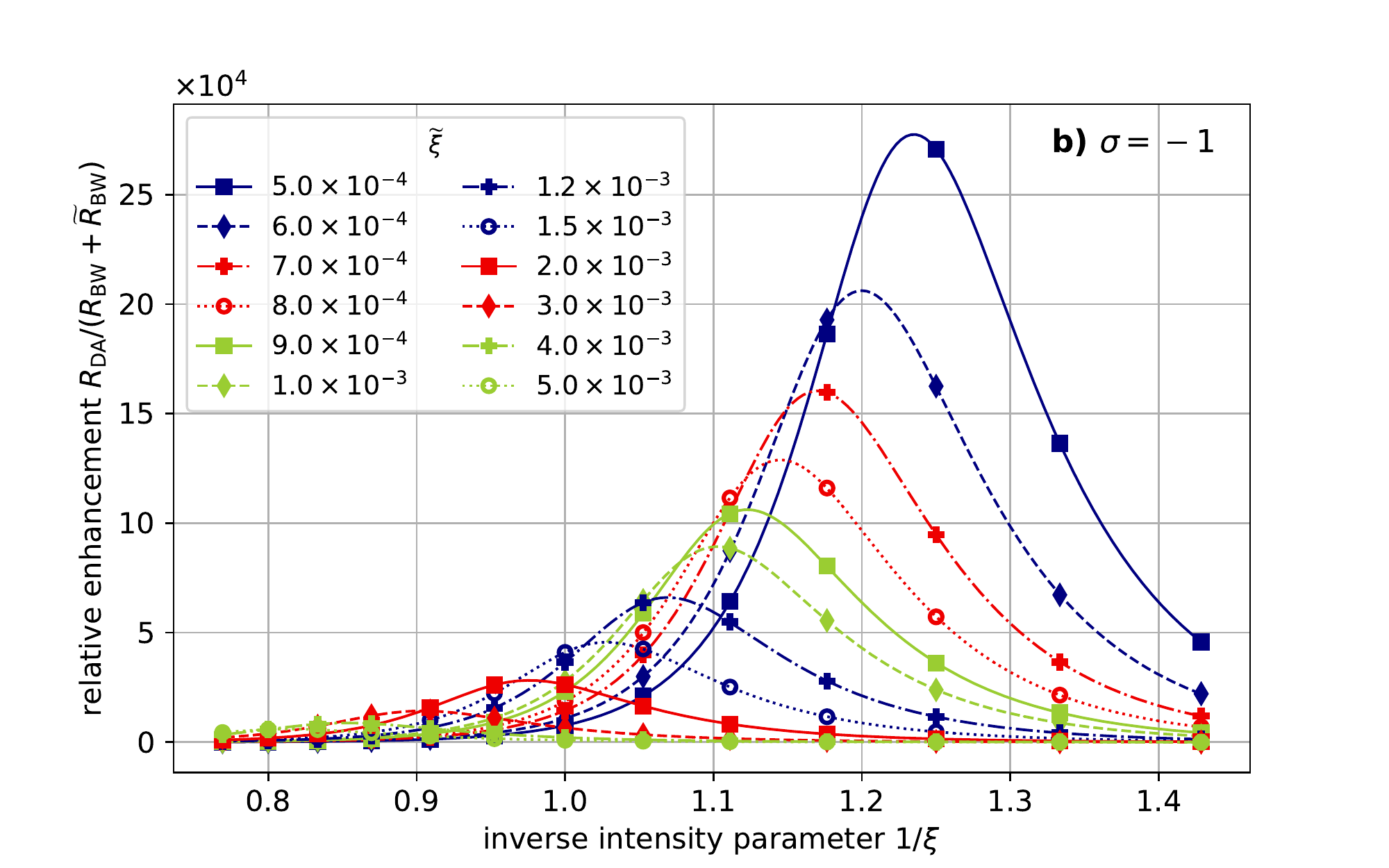}
\end{center}
\caption{Relative rate enhancement due to dynamical assistance for different values of the 
weak mode intensity parameter. The $\gamma$-photon and laser frequencies are the same as
in Fig.~\ref{fig:number}. Panels a) and b) refer to co- and counter-rotating laser modes,
respectively.}
\label{fig:enhancement-tilde-xi}
\end{figure}

\begin{figure}[t]
\begin{center}
\includegraphics[width=0.52\textwidth]{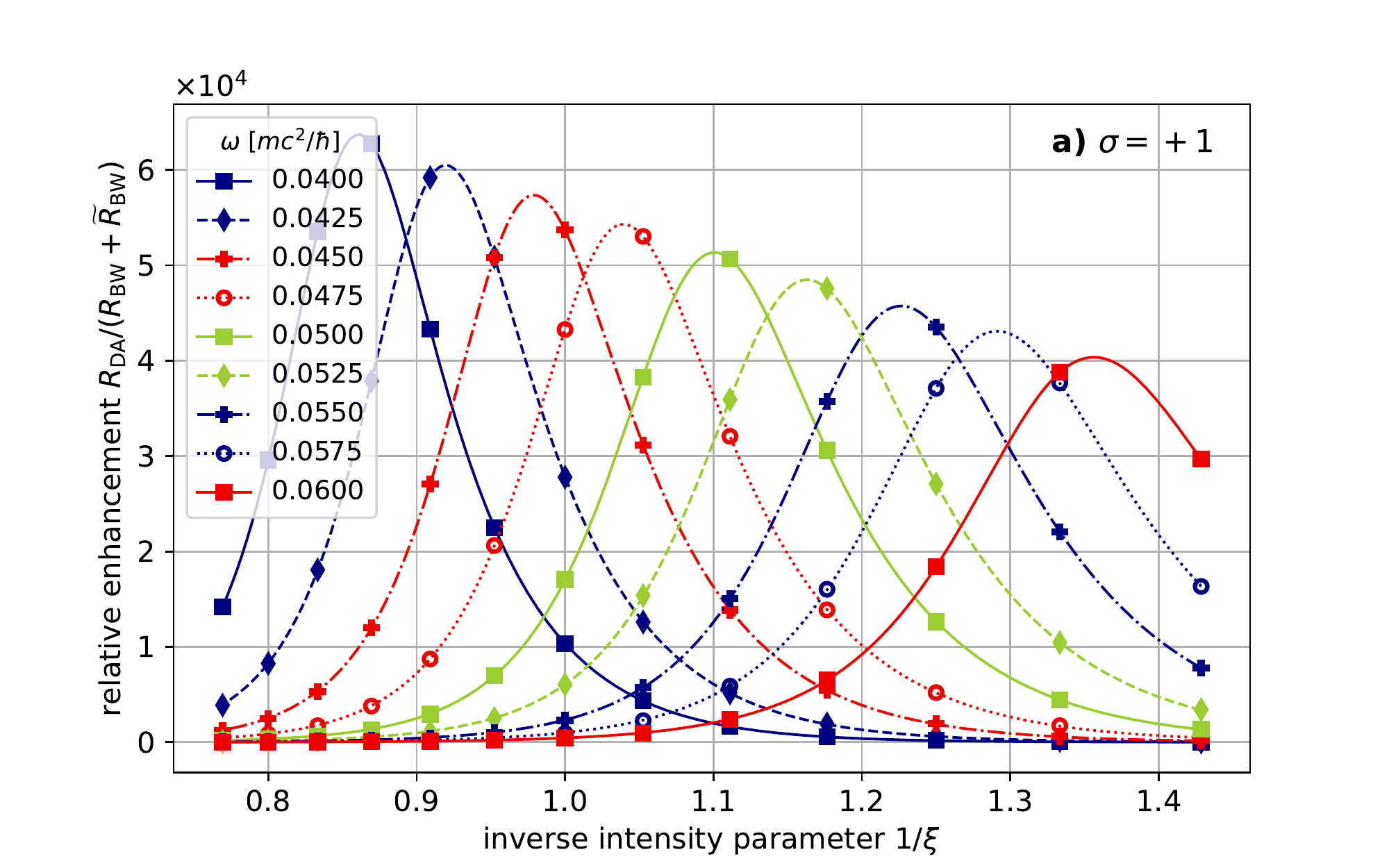}
\includegraphics[width=0.52\textwidth]{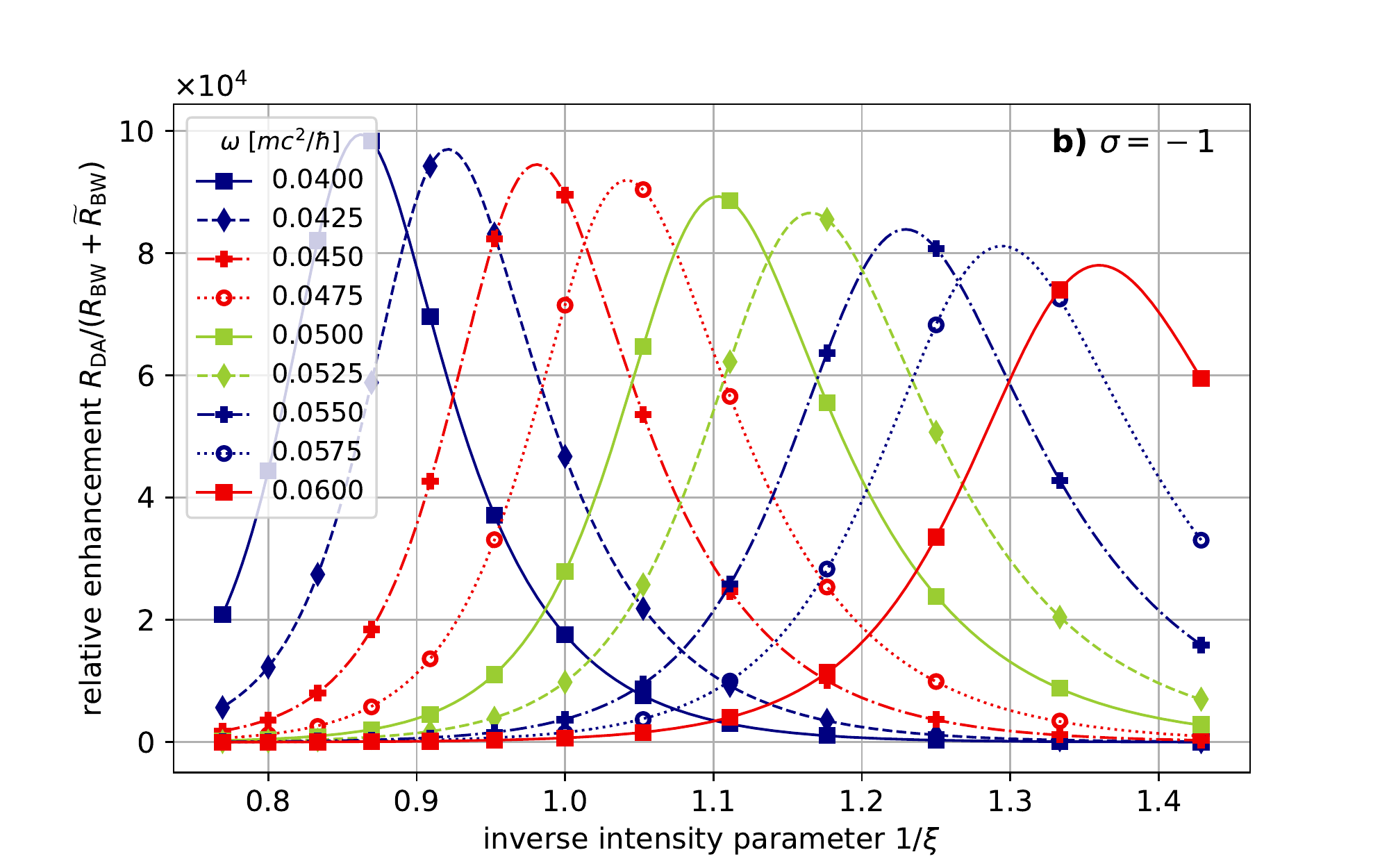}
\caption{Relative rate enhancement due to dynamical assistance for different values of the 
main mode frequency. The $\gamma$-photon frequency and assisting mode parameters are the same as
in Fig.~\ref{fig:number}. Panels a) and b) refer to co- and counter-rotating laser modes,
respectively.}
\label{fig:enhancement-omega}
\end{center}
\end{figure}

Figure~\ref{fig:enhancement-tilde-xi} shows the relative enhancement for various 
values of the weak mode intensity parameter in the interval $5\times 10^{-4}\le 
\tilde{\xi}\le 5\times 10^{-3}$. The dashed green lines with diamond symbols refer
to $\tilde{\xi}=10^{-3}$, as considered in Sec.~III.A. One sees that the maxima
of the relative enhancement curves shift to larger values of $1/\xi$ and increase
in magnitude, the smaller $\tilde{\xi}$ is. These trends can be understood by 
taking reference to Fig.~\ref{fig:inverse-xi} and noting that the dominant contributions
$R^{(1,-)}$ to the dynamically assisted rates $R_{\rm DA}(\sigma)$ scale with $\tilde{\xi}^2$, 
while the monochromatic rate $\tilde{R}_{\rm BW}$ scales more strongly with $\tilde{\xi}^6$; 
the unassisted rate $R_{\rm BW}$ remains unaltered when $\tilde{\xi}$ is varied.
Accordingly, when the value of $\tilde{\xi}$ is reduced, the rate $\tilde{R}_{\rm BW}$ 
decreases much more strongly than the rates $R_{\rm DA}(\sigma)$, while $R_{\rm BW}$ stays
the same (see Fig.~\ref{fig:inverse-xi}). The position of the maximum relative enhancement
thus shifts to the right towards larger values of $1/\xi$ and grows in magnitude.
We note, moreover, that the $\tilde{\xi}$-dependence for co-rotating and counter-rotating laser modes in
panels a) and b) of Fig.~\ref{fig:enhancement-tilde-xi} has very similar appearance.

In Fig.~\ref{fig:enhancement-omega}, the relative enhancement is displayed 
when the main mode frequency is varied in the range $0.04m\le\omega\le 0.06m$. The 
maxima of the curves decrease and move towards larger values of $1/\xi$ when $\omega$ grows.
That the relative enhancement reaches higher values when $\omega$ is small, can 
be attributed to the growing ratio of $\tilde{\omega}/\omega$ so that a single 
high-frequency photon corresponds to an increasing number of low-frequency photons.
Interestingly, the product $\omega\xi$ attains an approximately constant value at
all curve maxima, corresponding to an electric field strength of the main mode of 
$E\approx 0.045E_c$. Thus, close to this field strength, the dynamical assistance is
most efficient in the present scenario.

The influence of the main mode frequency is more pronounced for co-rotating laser modes 
[see Fig.~\ref{fig:enhancement-omega}\,a)]. The maximum relative enhancement decreases
from left to right by about 40\%, reaching $6.4\times 10^4$ for $\omega=0.04m$ and 
$4.0\times 10^4$ for $\omega=0.06m$. In contrast, the relative enhancement for 
counter-propagating laser modes in panel b) decrease only by about 20\%, from nearly
$10^5$ for $\omega=0.04m$ to $7.8\times 10^4$ for $\omega=0.06m$.

\begin{figure}[t]
\begin{center}
\includegraphics[width=0.52\textwidth]{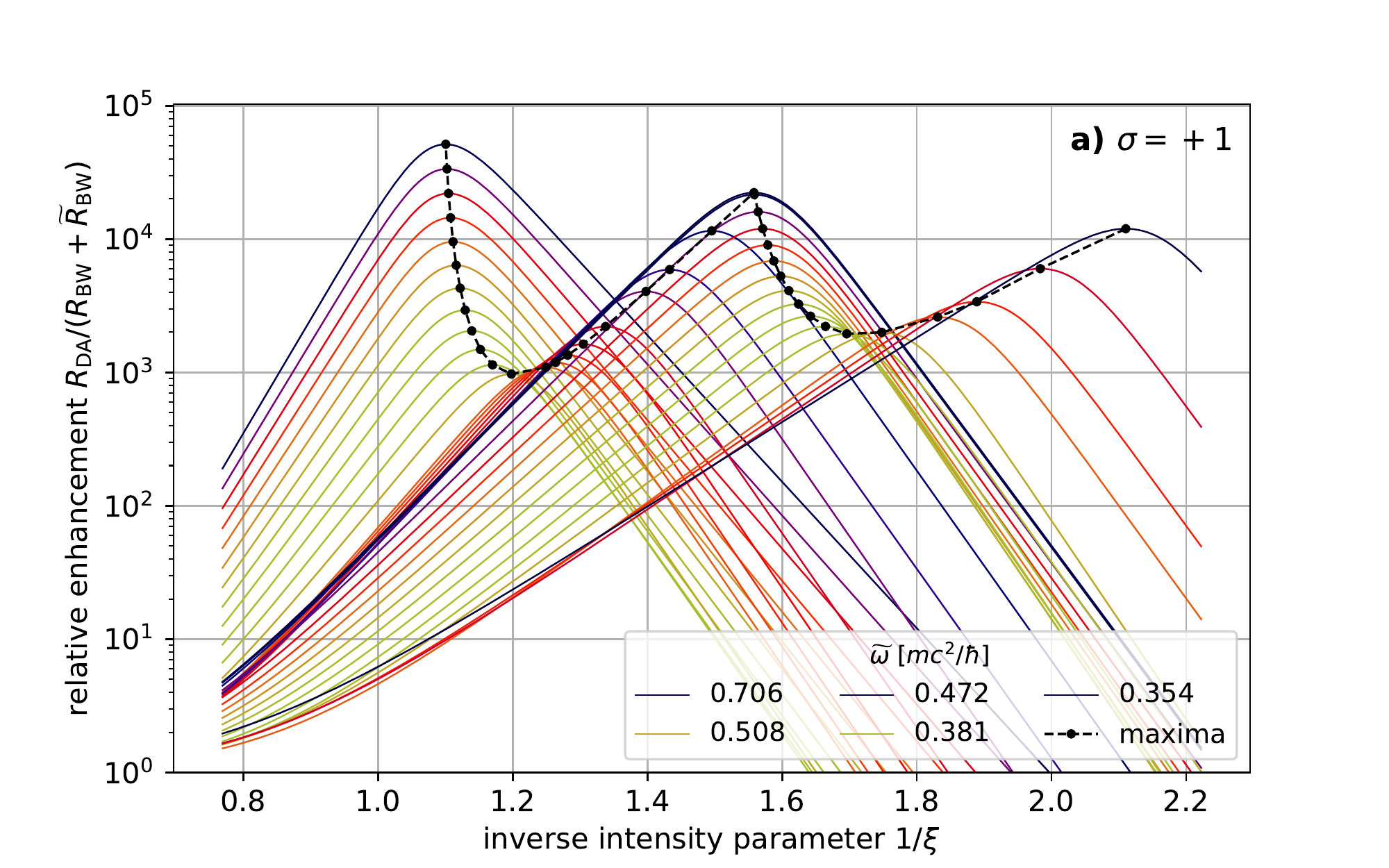}
\includegraphics[width=0.52\textwidth]{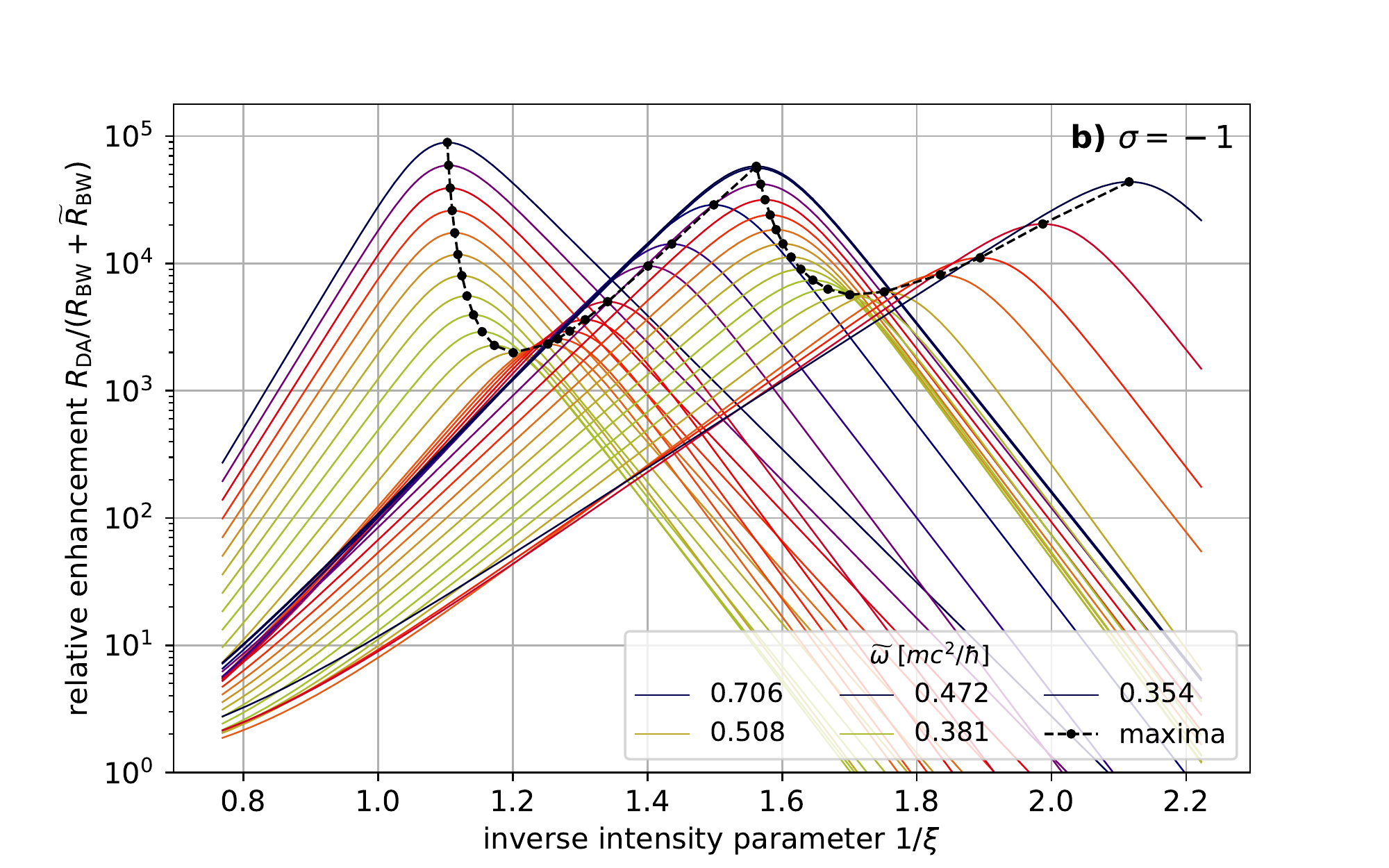}
\caption{Relative rate enhancement due to dynamical assistance for a sequence of decreasing
values of the weak mode frequency between $\tilde{\omega} = 0.706 m$ and $\tilde{\omega} = 0.354 m$ 
(from left to right). The $\gamma$-photon frequency, main mode frequency and assisting 
mode intensity parameter are the same as in Fig.~\ref{fig:number}. Black circles mark the 
curve maxima. Panels a) and b) refer to co- and counter-rotating laser modes, respectively.}
\label{fig:enhancement-tilde-omega}
\end{center}
\end{figure}

Very interesting structures arise when the relative enhancement is considered under 
variation of the assisting mode frequency $\tilde{\omega}$. The results are shown in 
Fig.~\ref{fig:enhancement-tilde-omega} within the interval $0.706m\ge\tilde{\omega}
\ge 0.354m$. Note that the figure legend contains selected $\tilde{\omega}$-values
in order to not overload it. The top curve on the left corresponds to the frequency
$\tilde{\omega}=0.706m$ that has been considered so far. When this value is lowered,
the relative enhancement curves go down and slightly shift to the right until 
$\tilde{\omega}\approx 0.5m$ is reached. Their decrease is due to the fact that the 
reduction of the tunneling barrier is less and less pronounced when the assisting mode 
frequency becomes smaller. 

However, when $\tilde{\omega}$ is decreased further, the relative enhancement curves
start to grow again and shift considerable to the right, until the value 
$\tilde{\omega}=0.472m$ is reached. At this frequency, the monochromatic process
of nonlinear Breit-Wheeler pair creation by the $\gamma$-beam and the assisting 
mode alone changes its character in the sense that now at least four (rather than
three) $\tilde{\omega}$-photons are required to overcome the creation threshold.  
The corresponding rate $\tilde{R}_{\rm BW}$ therefore drops down considerably, 
being suppressed by an additional factor of $\tilde{\xi}^2\ll 1$. As a consequence,
the crossing point with the unassisted rate $R_{\rm BW}$ shifts to the right
(see Fig.~\ref{fig:inverse-xi}), close to which the maximum relative enhancement
occurs, in agreement with the shift of the curve maxima arising in 
Fig.~\ref{fig:enhancement-tilde-omega}. 
When the assisting mode frequency is even further decreased, the same transition
takes place again. The curves decrease until $\tilde{\omega}\approx 0.38m$ is 
reached, afterwards start to increase again up to the value $\tilde{\omega}=0.354m$,
from which on at least five $\tilde{\omega}$-photons must be absorbed to produce pairs
via the monochromatic channel associated with the rate $\tilde{R}_{\rm BW}$. 

In order to highlight the gradual transitioning due to multiphoton channel 
closings, we have marked the maxima of the relative enhancement curves in 
Fig.~\ref{fig:enhancement-tilde-omega} by black circles.
The distinguished frequencies of the assisting mode $\tilde{\omega}\in\{0.706m, 0.472m, 
0.354m \}$, where the main maxima in Fig.~\ref{fig:enhancement-tilde-omega} arise, 
are of the form $\tilde{\omega}\approx \sqrt{2}m/\tilde{n}_0$ with 
$\tilde{n}_0\in\{2,3,4\}$, corresponding to $\tilde{s} = 4\tilde{\omega}\omega'
\approx 4m^2/\tilde{n}_0$.

\section{Conclusion}
Dynamically assisted nonlinear Breit-Wheeler pair creation has been studied in a 
bichromatic laser field, consisting of a strong main mode of low frequency and a 
weak assisting mode of high frequency, both with circular polarization. By taking 
the weak mode in leading order into account, an integral expression for the 
corresponding pair creation rate with absorption of one high-frequency laser photon 
has been obtained, whose structure resembles the well-known formula for the ordinary 
(i.e.~unassisted) nonlinear Breit-Wheeler process \cite{Reiss-1962, Nikishov-Ritus, 
Greiner}. 

We have shown that the assistance from the weak laser mode can enhance the pair 
creation rate by several orders of magnitude. In particular, the enhancement is 
more pronounced when the two laser modes have opposite helicities. This 
interesting effect can be attributed to a lowering of the angular momentum barrier 
of the process. The relative enhancement due to dynamical assistance, as compared 
with the rates for nonlinear Breit-Wheeler pair creation by each laser mode 
separately, was found to be the larger, the smaller the main mode frequency and 
the assisting mode intensity parameter are. Quite complex structures have arisen 
when instead the assisting mode frequency is varied, which are caused by 
multiphoton channel closings.

Our analysis complements a previous study where dynamically assisted nonlinear 
Breit-Wheeler pair creation of scalar particles was considered for mutually orthogonal, 
linearly polarized laser waves \cite{Jansen-PRA2013}. It is of potential relevance for 
future experiments on strong-field pair creation at high-intensity laser facilities 
\cite{ELI, E320, CALA, LUXE, RAL}.

\section*{Acknowledgment}
This work has been funded by the Deutsche Forschungsgemeinschaft (DFG) 
under Grant No. 392856280 within the Research Unit FOR 2783/1.



\begin{thebibliography}{33}
\bibitem{asymptotic} In accordance with $R_{\rm BW}\sim \exp[-8/(3\chi)]$,
the asymptotic value for $\xi\gg 1$ would be $b_0\approx 37.8$.
\bibitem{semiclassical} The semiclassical picture could be corroborated by redoing 
our $S$-matrix calculations in a spin-resolved manner (similarly to Refs.~\cite{spin1,
spin2,spin3}, for instance) which, however, is beyond the scope of the present study.
\end{thebibliography}
\end{document}